\documentclass[twocolumn,pra,showpacs,aps,superscriptaddress]{revtex4}

\usepackage{amsmath}
\usepackage{amsbsy}
\usepackage{amsthm}
\usepackage{amssymb}
\usepackage{latexsym}
\usepackage[dvips]{color,graphicx}

\begin{document}

\title{Collapse and revival oscillations as a probe\\
for the tunneling amplitude in an ultra-cold Bose gas }

\author{F. Alexander Wolf}
\affiliation{Department of Physics, Georgetown University, Washington, DC 20057, USA}
\affiliation{Theoretical Physics III, Center for Electronic Correlations and Magnetism,
Institute of Physics, Augsburg University, 86135 Augsburg, Germany}
\author{Itay Hen}
\affiliation{Department of Physics, Georgetown University, Washington, DC 20057, USA}
\author{Marcos Rigol}
\affiliation{Department of Physics, Georgetown University, Washington, DC 20057, USA}

\begin{abstract}
We present a theoretical study of the quantum corrections to the revival time 
due to finite tunneling in the collapse and revival of matter wave interference after
a quantum quench. We study hard-core bosons in a superlattice potential and the 
Bose-Hubbard model by means of exact numerical approaches and mean-field theory.
We consider systems without and with a trapping potential present. We show
that the quantum corrections to the revival time can be used to accurately 
determine the value of the hopping parameter in experiments with ultracold bosons in
optical lattices.
\end{abstract}
\pacs{03.75.Kk, 03.75.Hh, 05.30.Jp, 02.30.Ik}
\maketitle

\section{Introduction}

Collapse and revival oscillations in a Bose-Einstein condensate loaded in an optical
lattice were first experimentally observed in 2002 \cite{greiner02} and since then 
have been a subject of much theoretical and experimental interest \cite{bloch08}.
This phenomenon is understood as an oscillation between an initial coherent state and 
a final non-coherent (collapsed) state in a lattice where, after a quench, the hopping 
parameter between sites is negligible. Very recently it has been argued that such 
collapse and revival oscillations can be used as a very sensitive probe for effective 
three-body and higher interactions \cite{will09} by studying the time evolution of the 
visibility of the interference pattern. This has been investigated theoretically 
in Ref.~\cite{johnson09}. It was assumed both in experiment and in theory \cite{will09,johnson09}
that the initial state is a coherent state and that after the quench 
the tunneling amplitude is negligible, that is, that the systems were in the atomic limit. 
Following these assumptions, the time-evolving state is a product of coherent states 
localized at each lattice site and one deals with an effective one-site problem.

Our goal in this article is to go beyond the previous analysis and present a full 
many-body study of collapse and revival phenomena in one-, two-, and three-dimensional cubic lattices. 
We study hard-core bosons in the presence of a superlattice 
and the Bose-Hubbard model. For both cases we consider systems without and with a 
trapping potential present. For the hard-core case we use exact numerical approaches 
in one and two dimensions and compare them with the predictions of a Gutzwiller 
mean-field theory. We show that the latter is qualitatively and quantitatively 
correct when determining the revival time for small hopping amplitudes. Building on that, 
we present an analysis for the Bose-Hubbard model that is solely based on the Gutzwiller 
mean-field approach. We also provide an analytical solution for the homogeneous 
hard-core case.

A previous study \cite{fischer08} considered the effect of a finite hopping on 
the damping of the collapse and revival oscillations in a lattice without a confining
potential. The quantitative results presented there applied to an initial coherent state 
as in the articles mentioned previously. In contrast, we study the 
dynamics starting from initial states that are the exact many-body ground state
in some cases and the appropriate Gutzwiller ansatz in the other cases.

We find the functional form of those corrections to the revival time in the atomic limit, 
and show that if one knows the values of the superlattice potential for the 
hard-core case or the onsite interaction $U$ for the Bose-Hubbard model, 
such corrections can be used to accurately determine the tunneling amplitudes 
in experiments in optical lattices. In the atomic limit, for the Bose-Hubbard model, 
the revival time $t_\text{rev}$ is $t^\text{atom}_\text{rev}=2\pi\hbar/U$ 
\cite{wright96,imamoglu97}. Our general strategy is to numerically calculate 
the deviations from this value for $0<J/U<1$. Given that in experiments with ultracold 
gases in optical lattices the hopping parameter $J$ is exponentially sensitive to the
lattice depth, while the onsite repulsion $U$ is only power-law dependent on the lattice 
depth, this method of determining $J$ by means of the revival time may be more accurate 
than the approaches followed so far. Our results are also relevant to cases in which 
the tunneling is small but one is still interested in estimating its value.

The article is organized as follows. In Sec.\ \ref{exact}, we study the collapse 
and revival in the hard-core case in the presence of a superlattice potential.
This is done using numerically exact methods in one and two dimensions. 
In Sec.\ \ref{meanfield}, we introduce the time-dependent mean-field approach to the 
hard-core boson problem and compare its results with numerically exact ones in order 
to quantify the predictive power of the mean-field approximation for the revival time. 
In addition, we make some general theoretical statements and provide an analytical 
solution for the case without the trapping potential. Section \ref{soft-core} is devoted 
to analyzing the Bose-Hubbard model in three dimensional systems of soft-core bosons
with and without confining potentials present. 
Finally, we present our conclusions in Sec.\ \ref{conclusions}.

\section{Numerically exact results for hard-core bosons in a superlattice}\label{exact}

We first introduce the model Hamiltonian considered to study hard-core bosons in a 
superlattice. We also discuss its relation 
to the interaction quench in the soft-core case that is studied in Sec.\ \ref{soft-core}. 

\subsection{Motivation and methods}

The hard-core boson Hamiltonian on a superlattice with period $2$ in the presence of a harmonic
confining potential can be written as
\begin{equation} \label{hcb} 
\hat{H}_\text{HCB} = - J \sum_{\langle i j \rangle} 
\big( \hat{b}^\dagger_{i} \hat{b}^{}_{j} + \mathrm{H.c.} \big) 
+ A \sum_i (-1)^i \hat{n}_i 
+ V \sum_i r_i^2 \hat{n}_i \,,
\end{equation}
where the hard-core creation and annihilation operators at site $i$ are denoted by 
$\hat{b}^{\dagger}_{i}$ and $\hat{b}^{}_{i}$, respectively, and the local density 
operator by $\hat{n}_i=\hat{b}^{\dagger}_{i}\hat{b}^{}_{i}$. Hard-core boson operators
satisfy standard commutation relations for bosons. However, on the same site, they also 
satisfy the constraint $\hat{b}^{\dagger 2}_{i}= \hat{b}^2_{i}=0$, which precludes 
multiple occupancy of the lattice sites. The other parameters in Eq.\ (\ref{hcb}) 
are the hopping constant $J$ between nearest-neighbor sites $\langle i j \rangle$, 
the strength of the superlattice potential $A$, and the curvature of the harmonic trap $V$. 
The distance from site $i$ to the center of the trap $r_i$ is measured in units of the 
lattice constant $a$, which we set to unity. In what follows, we also denote
the total number of lattice sites by $L$ and the total number of particles by $N$.

The hard-core model on a superlattice is particularly suitable to study collapse
and revival phenomena because in one dimension (1D) it can be exactly solved by means
of the Jordan-Wigner mapping to noninteracting fermions \cite{jordan28}. In equilibrium,
these systems were studied in Ref.\ \cite{rousseau06}, where they were shown 
to exhibit ground state phases that are similar to those of the Bose-Hubbard model. The 
nonequilibrium dynamics of hard-core bosons in a superlattice potential was
studied in Ref.\ \cite{rigol06}, where collapse and revival oscillations of the momentum
distribution function were observed.

A quench of the superlattice potential $A$ in Eq.\ (\ref{hcb}) has a similar effect to
a quench of $U$ in the Bose-Hubbard model. From a simple band-structure calculation it 
follows that in 1D $A$ opens a gap $\Gamma=2A$ in the hard-core boson energy 
spectrum \cite{rousseau06,rigol06},
\begin{equation}
\epsilon_{\pm}(k) = \pm \sqrt{ 4 J^2 \cos^2(ka) + A^2} \,,
\end{equation}
where ``$+$'' (``$-$'') denotes the upper (lower) band. In two (2D) and three (3D) dimensions, 
hard-core bosons cannot be mapped to noninteracting fermions. The phase diagrams for 
the ground state of such systems were studied in detail in Refs.\ \cite{hen09,hen10} 
by various numerical and analytical approaches, and were shown to be qualitatively  
similar to the phase diagrams of the Bose-Hubbard model in 2D and 3D
\cite{fisher89,freericks96,sansone08}.  

As already noted, in the atomic limit of the Bose-Hubbard model, the revival time
after the interaction quench is given by $t^\text{atom}_\text{rev}=2\pi/U$ 
(we set $\hbar\equiv1$ henceforth); similarly, for hard-core bosons in a superlattice potential,
it follows that $t^\text{atom}_\text{rev}=\pi/A$ \cite{rigol06}. 

Hence, in this section we take advantage of the fact that the nonequilibrium dynamics of  
hard-core bosons in 1D can be exactly solved for large system sizes by means of the 
approach presented in Refs.~\cite{rigol05}, which makes use of the Jordan-Wigner 
transformation to noninteracting fermions \cite{jordan28}. In 2D, due to the reduced
Hilbert space (when compared to soft-core bosons), one can perform full diagonalization 
calculations for small, but meaningful, periodic systems. All our exact results for 2D 
hard-core systems were obtained in $4\times 4$ lattices with periodic boundary conditions.

The two preceding approaches allow us to make exact predictions for the quantum corrections 
due to finite hopping amplitudes to the revival time $t^\text{atom}_\text{rev}$
\begin{equation}
 \Delta t_\text{rev} = t^\text{atom}_\text{rev}-t_\text{rev},
\end{equation}
which in turn will help us gauge the accuracy of the mean-field approach that we use later 
for studying the Bose-Hubbard model.

In the latest experimental and theoretical studies \cite{will09,johnson09}, 
the main observable 
under consideration was the visibility of the interference pattern. 
Here, instead, we focus
our attention on the time evolution of the $n_{k=0}$ momentum occupation number,
\begin{equation}
n_{k=0}=\dfrac{1}{L}\sum_{ij} \langle\hat{b}^\dagger_{i} \hat{b}^{}_{j} \rangle,
\end{equation}
which is also measured in experiments.

\subsection{Results for hard-core bosons in a periodic potential in 1D and 2D}\label{periodic}

In all cases for hard-core bosons presented here we consider the following quench.
A system is prepared in a superfluid state with $J_\text{ini}=1$ (which sets our energy scale) 
and no superlattice potential. At time $t=0$ the 
superlattice potential $A$ is quenched to a constant value $A_\text{fin}=1$
and the hopping constant $J$ is reduced to several values $J_\text{fin}<1$
(in the remainder of the text,
the notation $A_\text{fin}\equiv A$ and $J_\text{fin}\equiv J$ is used
in all unambiguous cases).

\begin{figure}[!htb]
\begin{center}
\includegraphics[width=0.41\textwidth]
{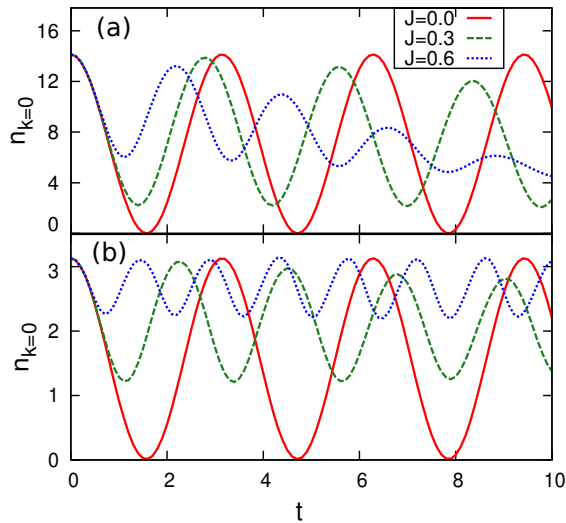}
\end{center} \vspace{-0.7cm}
\caption{(Color online) 
Time evolution of $n_{k=0}(t)$ for three final values for the hopping constant,
$J_{\text{fin}}=0$, $J_{\text{fin}}=0.3$, and $J_{\text{fin}}=0.6$, 
in a system with a superlattice potential in 1D (a)
and 2D (b).
These calculations were done in a chain with $L=400$ lattice sites 
in 1D and in a $L=4\times4$ system in 2D,
both at quarter filling.  
Time is measured in units of $\hbar / J_{\text{ini}}$.}
\label{exactHomEvol}
\end{figure}

In Fig.\ \ref{exactHomEvol}, we show the time evolution of $n_{k=0}$ after this quench
in (a) a chain and (b) a $4\times 4$ cluster, both at quarter filling. 
Results are presented for three different final values of $J$, where the atomic limit ($J=0$) 
revival time can be clearly seen to correspond with the prediction 
$t^\text{atom}_\text{rev}=\pi/A$. Two effects of finite final hopping are evident in those 
plots: first, a clear shift in the frequency of the oscillations
and, second, a damping of the amplitude. In the following, we restrict our analysis
to the period and amplitude of the first revival. In the homogeneous case, the frequency
can be calculated from the revival time. In the presence of a confining potential, 
this is in general not possible as the exact revival time can change on long time scales.

\begin{figure}[!htb]
\begin{center}
\includegraphics[width=0.41\textwidth]
{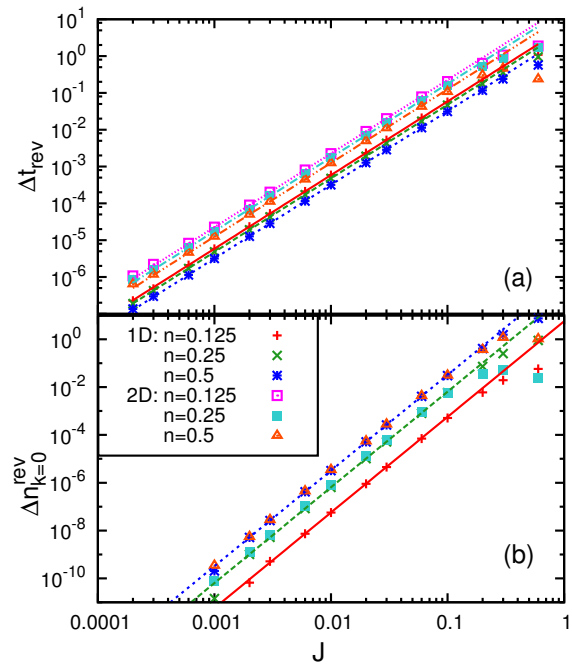}
\end{center} \vspace{-0.7cm}
\caption{(Color online)
Quantum corrections to (a) the revival time ($\Delta t_\text{rev}$), and 
(b) the revival amplitude ($\Delta n_{k=0}^\text{rev}$), vs $J$. Results are presented 
for three densities $n=0.125$, $n=0.25$ and $n=0.5$ in 1D and 2D. In (a) a quadratic 
dependence is observed whereas in (b) a quartic dependence is present. These behaviors are 
emphasized by power-law fits for data points with $J\leq0.01$. The system 
sizes for 1D are $L=800$ for $n=0.125$, and $L=400$ for $n=0.25$ and $n=0.5$, and for 2D
$L=4\times4$ for all densities. Results for $n>0.5$ trivially follow from the particle-hole
symmetry of the model. No data are presented for $n=0.125$ in (b) because only 
two particles are present in the $4\times4$ cluster and no damping occurs.}
\label{exactHom}
\end{figure}

The quantum corrections due to finite hopping $\Delta t_\text{rev}$ versus 
$J$ for constant $A=1$ are presented in a log-log plot in Fig.~\ref{exactHom}(a) for 1D and 2D systems. 
We find from those plots that the corrections follow a quadratic behavior 
for values of $J\lesssim 0.1$. This is depicted by a quadratic fit $f(x)=a x^2$ to 
data points with $J\leq0.01$. Also evident from those plots is the very weak dependence
of $\Delta t_\text{rev}$ on the density both in 1D and 2D. This turns 
out to be very convenient later when studying the harmonically trapped systems.

In Fig.~\ref{exactHom}(b) we consider the damping of the oscillations, which can be 
characterized by the amplitude of the first revival 
$n_{k=0}^\text{rev}=n_{k=0}(t=t_\text{rev})$ subtracted from its value in the atomic limit:
$\Delta n_{k=0}^\text{rev}=n_{k=0}^\text{atom,rev}-n_{k=0}^\text{rev}$. 
For this quantity we find a quartic behavior, as illustrated by the fits
in Fig.\ \ref{exactHom}(b) and a much stronger dependence on 
the density. The very fast reduction of
the damping with decreasing $J$ makes it a less attractive tool for experimentally
probing small values of $J$.

We find numerically the scaling of $t_\text{rev}$ with respect to the system 
parameters $J$ and $A$ to have the 
following functional form $t_\text{rev}(J,A)\equiv t_\text{rev}(J/A)/A$ 
whereas for the damping $n_{k=0}^\text{rev}(J,A)\equiv n_{k=0}^\text{rev}(J/A)$; that is, 
the former depends on
the value of $A$ and $J/A$ while the latter is only a function of the ratio $J/A$.
In the atomic limit, the revival time scales with $A$: $t_\text{rev}^\text{atom}(A)= \pi/A$ and 
therefore the preceding scaling holds also true for 
$\Delta t_\text{rev}$: $\Delta t_\text{rev}(t,A)\equiv  \Delta t_\text{rev}(t/A)/A$.
In Sec.\ \ref{meanfield} we are able to analytically confirm this
result for the mean-field approximation. On the other hand, in the atomic limit, 
$n_{k=0}^\text{max}(A)$ is independent of $A$ as the system exhibits perfect revivals so
$\Delta n_{k=0}^\text{rev}$ is only a function of $J/A$.

\subsection{Results for hard-core bosons in a trap in 1D}\label{sec:hcbtrap}

Experimental systems are in general different from the ones discussed in 
Sec.\ \ref{periodic}. This is because a confining potential is required for containing 
the gas of bosons. 
The confining potential in experiments is to a good approximation harmonic, 
and generates an inhomogeneous density profile. Given the results shown in 
Fig.\ \ref{exactHom}(a), where the revival time was shown to depend only weakly on the density, 
one would expect the outcome in the presence of a trap not to be strongly dependent on the confining 
potential and the total number of particles.

\begin{figure}[!htb]
\begin{center}
\includegraphics[width=0.41\textwidth]
{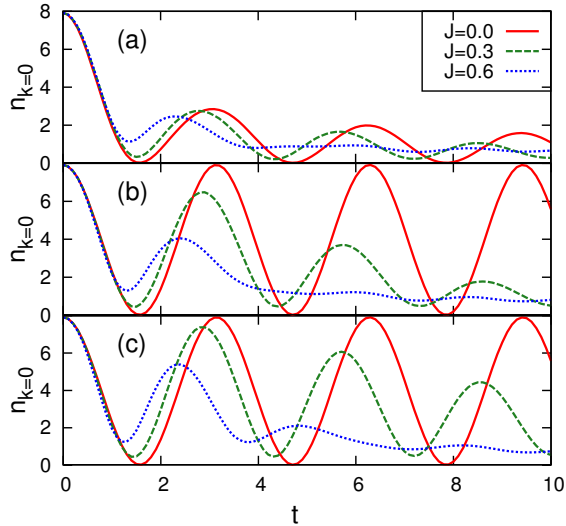}
\end{center} \vspace{-0.7cm}
\caption{(Color online) 
Time evolution of $n_{k=0}(t)$ in a trap after a quench with 
(a) constant curvature $V=10^{-4}$, 
(b) constant characteristic density $\tilde\rho = 1.0$, 
and (c) turning off the trap ($V_\text{fin}=0$), all for the same initial state. 
In text (b) and (c) are referred to 
as quench type (i) and (ii), respectively.
These calculations were done for a system with $L=400$ and $N=100$. 
To keep the characteristic density constant 
during the quench we changed $V_\text{ini}=10^{-4}\rightarrow V_\text{fin}=0$ for $J=0$, 
$V_\text{ini}=10^{-4}\rightarrow V_\text{fin}=3\times 10^{-5}$ for $J=0.3$ and 
$V_\text{ini}=10^{-4}\rightarrow V_\text{fin}=6\times 10^{-5}$ for $J=0.6$.}
\label{exactTrapEvol}
\end{figure}

Up to small corrections, the preceding turns out to be the case for the changes 
induced in the revival time by a finite hopping. However, as shown in 
Fig.\ \ref{exactTrapEvol}(a), if one quenches $J$ and $A$ keeping constant the trapping 
potential, then a very high damping rate can be seen even in the atomic limit. Hence, 
measurements at a constant curvature of the trap are not the best way to proceed in 
trapped systems. They mix the effects of the trapping potential and the finite hopping 
in the outcome. In fact, even the quadratic behavior obvious in the homogeneous 
case (Fig.~\ref{exactHom}) becomes obscured in the trap if the confining potential
is kept the same from the initial state.

In previous work in equilibrium it has been argued that the correct way to define
the thermodynamic limit for a trapped system is by keeping constant the so-called
characteristic density $\tilde{\rho} = N [V/(dJ)]^{\frac{d}{2}}$,
where $d$ is the dimensionality of the system (see, e.g., Ref.\ \cite{rigol09}).
This is equivalent to what is done in homogeneous systems when one keeps constant the
density $N/L$. Since in Sec.\ \ref{periodic} all quenches were performed keeping constant 
$N/L$, we have studied quenches in the trap in which the characteristic density is kept 
constant; that is, one needs to reduce the trapping potential by the same amount that 
the hopping parameter is reduced. We denote this scenario quench type (i).
Another way of reproducing the homogeneous results that comes to mind is to remove
the trapping potential concurrently with the superlattice quench and observe 
oscillations which then take place in a homogeneous potential. 
This scenario is denoted quench type (ii). Within the
second approach, one realizes that the gas starts expanding after the quench. 
However, if the considered hopping parameters and revival times are sufficiently small, 
this will not constitute a problem. Results for the dynamics of these cases
are shown in Figs.\ \ref{exactTrapEvol}(b) and \ref{exactTrapEvol}(c). 
In contrast to the quench that keeps constant the curvature of the trap
we now observe that the time evolution of $n_{k=0}$ is very similar 
to the one in homogeneous systems depicted in Fig.\ \ref{exactHomEvol}.

\begin{figure}[!htb]
\begin{center}
\includegraphics[width=0.41\textwidth]
{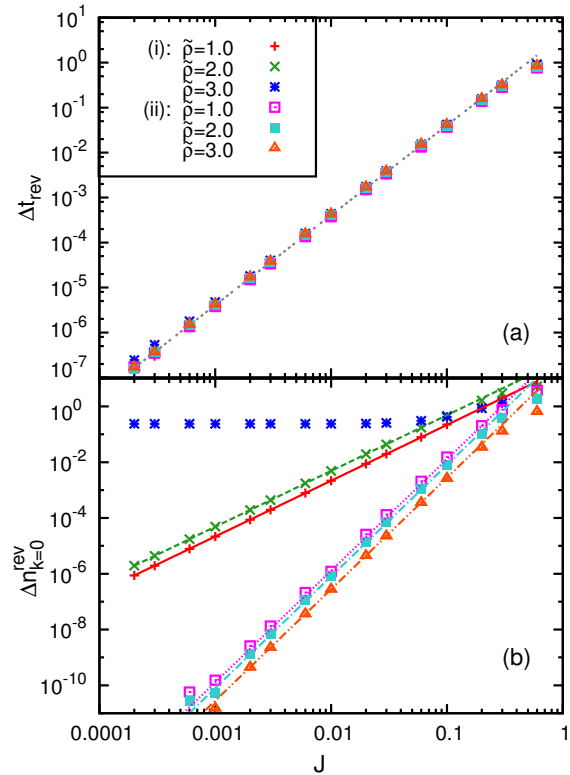}
\end{center} \vspace{-0.7cm}
\caption{(Color online) 
Quantum corrections to (a) the revival time ($\Delta t_\text{rev}$), and 
(b) the revival amplitude ($\Delta n_{k=0}^\text{rev}$) vs $J$ for $A_\text{fin}=1$, both
for the quench scenarios (i) keeping the characteristic density constant,
and (ii) turning off the trap. Results are
presented for three different initial values of the characteristic density
$\tilde{\rho}=1.0$, $\tilde{\rho}=2.0$, and $\tilde{\rho}=3.0$, which correspond 
to curvatures $V_\text{ini}=4.44\times10^{-5}$, $V_\text{ini}=1.78\times10^{-4}$, 
and $V_\text{ini}=4\times10^{-4}$, 
respectively; the system sizes are $L=500$, $L=400$, $L=300$, with $N=150$
in all cases. Note that $\tilde\rho=3.0$ already has a Mott-insulating
region in the center of the trap, while $\tilde\rho=1.0$ and $\tilde\rho=2.0$ 
are entirely superfluid. The quadratic and quartic fits in (a) and (b) were done for points 
with $J\leq0.01$.}
\label{exactTrap}
\end{figure}

For both quench scenarios (i) and (ii), we find a quadratic
behavior for $\Delta t_\text{rev}$, which is similar to what was shown 
in Fig.~\ref{exactHom} for homogeneous systems. 
Interestingly this behavior is, as depicted in Fig.\ \ref{exactTrap}(a),
practically independent of the quench type and the 
characteristic density of the initial state. We note that for 
$\tilde{\rho}=3.0$ the initial state has an insulating ($n=1$) domain in the center
of the trap, while for the other characteristic densities the system is purely
superfluid. In 1D, the insulator appears
in the center of the trap when $\tilde{\rho}\sim 2.6-2.7$ \cite{rigol05}.
The independence of the asymptotic behavior of $\Delta t_\text{rev}$ on the initial state 
suggests that by measuring the 
correction to the revival time due to the finite value of $J$ in experiments,
it is possible to determine $J$ if one knows $A$. The same can be said for 
systems without a trap.

On the other hand, as shown in Fig.\ \ref{exactTrap}(b), $\Delta n_{k=0}^\text{rev}$  
reveals a strong dependence on the quench type and the initial density. For 
scenario (i), we obtain a quadratic behavior for pure superfluid initial states
($\tilde\rho=1.0$ and $\tilde\rho=2.0$), while for the one having a Mott insulating domain 
($\tilde\rho=3.0$) a constant damping rate is always present for $J<1$. 
For quench type (ii), the damping behaves completely differently and shows the 
quartic behavior observed in the homogeneous case. With respect to the aim to simulate the homogeneous
case, this result indicates that the fact that the density profile remains unchanged during
the time evolution under scenario (i) is less important than the fact that the potential 
is homogeneous after quench (ii).

\section{Mean-field approach}\label{meanfield}

Within the mean-field approximation, we can extend our analysis to consider the more
experimentally relevant Bose-Hubbard model:
\begin{multline}\label{scb}
\hat{H}_{\text{SCB}} 
= - J \sum_{ \langle i j \rangle}(\hat{a}^\dagger_{i} \hat{a}_{j} + \mathrm{H.\,c.} ) \\
+ \frac{U}{2} \sum_{i} \hat{n}_i  (\hat{n}_i-1)  
+ \sum_{i} \hat{n}_i V r_i^2  \,,
\end{multline}
where $[\hat{a}_i,\hat{a}_j^\dagger]=\delta_{ij}$ and
$[\hat{a}^{}_{i},\hat{a}^{}_{j}]=[\hat{a}^{\dagger}_{i},\hat{a}^{\dagger}_{j}]=0$,
as usual for bosons. The on-site interaction energy is denoted by $U$.

The mean-field theory that we employ is based on the restriction
of the wave function to the Gutzwiller-type product state,
\begin{equation}\label{productState}
 \vert \Psi_{\mathrm{MF}} \rangle 
 = \prod_{i=1}^{L} \sum_{n=0}^{n_\text{c}} \alpha_{in} \vert n \rangle_i
 \, ,
 \end{equation}
where $n_\text{c} \rightarrow \infty$ for thermodynamic systems,
$\vert n \rangle_i$ denotes a single-site Fock state for lattice 
site $i$ and the complex coefficients $\alpha_{in}$
allow for a time dependence.
For all numerical calculations, a finite cutoff $n_\text{c}$ is taken.

The mean-field ground state in equilibrium is found by minimization of
the energy expectation value,
\begin{equation}\label{EQMin}
\langle \Psi_{\mathrm{MF}} \vert  \hat{H}_{\text{SCB}} - \mu \hat{N} \vert \Psi_{\mathrm{MF}} \rangle ,
\end{equation}
where $\mu$ is the chemical potential and $\hat{N}$ counts the 
total number of particles. Hence, from here on we work on the grand-canonical ensemble.
To find the time-evolution of the mean-field approximated
system, we employ the time-dependent variational principle \cite{jaksch02}
that minimizes the expression
\begin{equation}\label{NEQMin}
\langle \Psi_{\mathrm{MF}} \vert  i \partial_t - \hat{H}_{\text{SCB}} + \mu \hat{N} \vert \Psi_{\mathrm{MF}} \rangle \,,
\end{equation}
and yields the following set of differential equations:
\begin{multline}\label{jakschEq}
 i \dot\alpha_{in} 
 = - J \sum_{ \, \langle j \rangle_i} 
\big ( \sqrt{n+1}\, \alpha_{i(n+1)} \Phi_j^* + \sqrt{n}\, \alpha_{i(n-1)} \Phi_j \big) \\
 + \alpha_{i n}\, n \left[\, \frac{U}{2}(n-1) + V r_i^2 - \mu  \right] \,.
\end{multline}
Here $\Phi_j=\langle a_j\rangle=\sum_{n=1}^{n_\text{c}}\sqrt{n}\,\alpha_{j(n-1)}^*\alpha_{jn}$, 
$\alpha_{i(-1)} = \alpha_{i(n_\text{c} + 1)} = 0$, and $\sum_{\, \langle j \rangle_i}$ 
denotes summation over all $j$ that are nearest neighbors of $i$. 
This is a set of $L\times n_\text{c}$ equations. The time evolution described by
Eq.\ \eqref{jakschEq} preserves normalization and the total particle number $N$. 
We solve the system numerically using a fourth-order Runge-Kutta method.
Self-consistency is guaranteed by monitoring the total energy, particle number, and
normalization.

At this point it is important to stress that
this mean-field approach is in principle an uncontrolled approximation.
We gauge its validity against 
our exact results in Sec.\ \ref{meanfieldB}. Before doing so, we present
an instructive analytical solution for the equations introduced previously in the 
hard-core limit and for a periodic potential ($V=0$).

\subsection{Analytical mean-field solution for hard-core bosons in a periodic potential}

In the case of hard-core bosons, it is possible to reduce the number of equations considerably 
and employ a parametrization for the $\alpha_{in}$ in Eq.\ \eqref{jakschEq} that 
preserves normalization and deals with real variables -- this is due to the equivalence
of hard-core bosons to $s=1/2$ spins, which leads to the following ansatz for the Gutzwiller 
wave function \cite{ma85}:
\begin{equation}\label{spinState}
 \vert \Psi_{\mathrm{HCB}} \rangle 
 = \prod_{i=1}^{L} \mathrm{e}^{i \chi_i} 
   \left( \sin\frac{\theta_i}{2} + \cos\frac{\theta_i}{2} \mathrm{e}^{i \phi_i} 
   a_i^\dagger \right) \vert 0 \rangle\, .
\end{equation}

If there is no trap in the system ($V=0$)
it is possible to use translational invariance to simplify the equations \eqref{jakschEq},
in which in the hard-core limit we again employ the superlattice quench introduced
before. This leads to a formal replacement of $U(n-1)/2$ by $A$ in \eqref{jakschEq}.
As all sites with the same potential must have the same properties 
and the system's wavefunction is a product of single-site states,
the $L$-site system can be reduced to an effective two-site problem (for two on-site potentials 
$\mu_{1/2} = \mu \pm A$) independent of the dimension. Then the ansatz 
Eq.~\eqref{spinState} yields the following form of Eq.~\eqref{jakschEq}:
\begin{subequations}\label{dEq2site}\begin{align} 
 &\dot \theta_1 = - 2\,d\,J \sin \theta_2 \sin \phi \,, \label{dEq1} \\
 &\dot \theta_2 = 2\,d\,J \sin \theta_1 \sin \phi \,,  \label{dEq2}\\
 &\dot \phi = 2 A - 2\,d\,J 
 (\sin \theta_2 \cot \theta_1 - \sin \theta_1 \cot \theta_2 ) \cos\phi\,, \label{dEq3}
\end{align}\end{subequations}
where $\phi \equiv \phi_1 - \phi_2$. Here it can be seen that dimensionality enters 
the equations only by a simple rescaling of the hopping parameter: $J \rightarrow d\,J$.

\begin{figure}[!htb]
\begin{center}
\includegraphics[width=0.41\textwidth]
{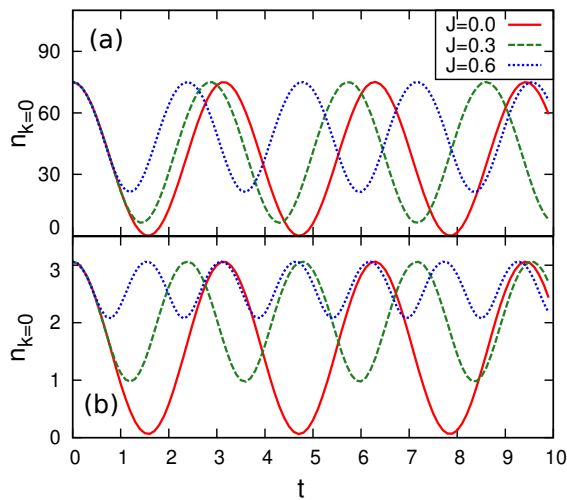}
\end{center} \vspace{-0.7cm}
\caption{(Color online) 
Plot of the time evolution of hard-core bosons in a periodic 
potential in the mean-field approximation for 1D (a) and 2D (b)
after a superlattice quench $A_\text{ini}=0\rightarrow A_\text{fin}=1$
for a system at quarter filling.
This is to be compared with the exact results in Fig.\ \ref{exactHomEvol},
where exactly the same system parameters were used.
}
\label{MFhomEvol}
\end{figure}

Also, the argument of translational invariance allows one to find a simple expression for $n_{k=0}$ in the two-site system:
\begin{equation}\label{nk0MF1}
n_{k=0} = n + \tfrac{1}{4}  \sin \theta_1 \sin \theta_2 \cos \phi\,.
\end{equation}
Figure \ref{MFhomEvol} depicts the time evolution of $n_{k=0}$ for the same systems for
which the exact solution was presented in Fig.\ \ref{exactHom}. One can clearly see that the
mean-field and exact results show a similar shift of the frequency. However, the mean-field
solutions exhibit no damping, and hence they are qualitatively incorrect for that quantity.

At this point it is instructive to extract the result for the atomic limit: for $J=0$, 
Eq.~\eqref{dEq2site} has the trivial solution $\phi(t) = 2A\,t + \phi(0)$
and $\theta_{1/2}$ constant. Insertion of this result in Eq.~\eqref{nk0MF1}
immediately reveals the revival time $t^\text{atom}_\text{rev}=\pi/A$.

Obtaining the solution of the system of Eqs.\ \eqref{dEq2site} for finite $J$
is possible by treating them like a classical system. Identification of Hamilton functions
and the observation of its trajectories leads to an analytical expression for the period of 
$n_{\mathrm{k=0}}$:
\begin{align}\label{period}
& t_\text{rev} = \int_{u_1}^{u_2} \mathrm{d}u \, f(u), \quad \text{where} \\
& f(u)
= \left\{ d^2J^2 (1-u^2) [1-(2 \gamma - u)^2]- (\mathcal{H}'_0 - A u)^2  \right\}^{-\frac{1}{2}} 
\nonumber
\end{align}
where $\gamma=2n-1$ and $\mathcal{H}_0'  =  - 4 n (1-n)\,d\,J -  \gamma A$.
A closed expression for the preceding integral exists. However, it is cumbersome and does not 
provide any apparent information on the functional form of $t_\text{rev}$ as it depends 
on the elliptic integral of the first kind. The integral limits $u_1$ and $u_2$ are the 
solutions of $1/f(u)=0$ that lie within $[-1,1]$. This requires solving the root of a polynomial 
of fourth order, which can also be done analytically. The lower boundary $u_1$ is given by a simple 
expression: $u_1 = \gamma$. In the case of half filling, also the upper boundary is given by a 
simple expression: $u_{2} = \frac{A}{2dJ}\left(\sqrt{1+8 d^2J^2/A^2} -1 \right)$.

\begin{figure}[!htb]
\begin{center}
\includegraphics[width=0.41\textwidth]
{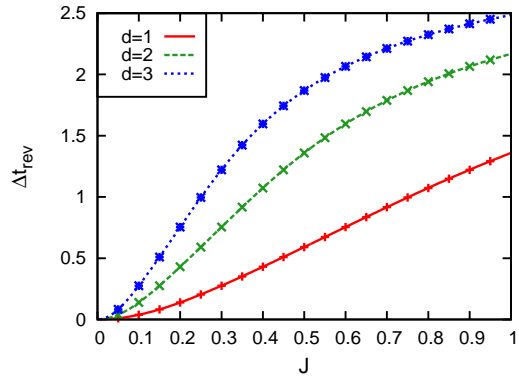}
\end{center} \vspace{-0.7cm}
\caption{(Color online) 
Plot of the revival time $\Delta t_\text{rev} = t_\text{rev}^\text{atom} - t_\text{rev}$ 
for a system of hard-core bosons in a periodic potential
and a superlattice quench $A_\text{ini}=0\rightarrow A_\text{fin}=1$. 
$t_\text{rev}$ is given by the analytic solution
Eq.\ \eqref{period}. Depicted are results for 1D, 2D, and 3D. The data points 
are numerical solutions of the set of Eqs.\ \eqref{jakschEq}.}
\label{analytic}
\end{figure}

In Fig.~\ref{analytic}, we plot the analytic solution for different dimensions at quarter filling. 
As mentioned before, dimensionality in the mean-field picture is captured by a simple rescaling 
of $J \rightarrow d\,J$. For comparison, we depict numerical solutions of Eq.\ \eqref{jakschEq}
as points in the plot. The latter are required for studying the inhomogeneous trapped 
hard-core boson case and soft-core bosons, for which no analytic solutions are available.

The analytic expression \eqref{period} allows us to confirm the numerical finding for the scaling
relations of $t_\text{rev}$ and $n_{k=0}^\text{rev}$ with respect
to the  parameters $A$ and $J$. The proposed scaling for the revival time
$t_\text{rev}(J,A)\equiv t_\text{rev}(J/A)/A$
does obviously hold for the integrand $f(u)$ in Eq.\ \eqref{period}. 
The fact that for the integration limits one has $u_{1/2}(J,A)\equiv u_{1/2}(J/A)$,
proves the proposition for the mean-field case.

Furthermore, the evaluation of Eq.\ \eqref{nk0MF1} leads to the expression
\begin{equation}\label{nk0MF2}
n_{k=0} = n - \tfrac{1}{4dJ} (\mathcal{H}_0' + A \cos \theta_1)\,,
\end{equation}
where one can see, (i) that no damping of $n_{k=0}$ occurs within the mean-field 
approximation in the hard-core limit and (ii) that $n_{k=0}$ only depends on $J/A$
as found for the numerical solution.

We note also that for the case of half filling and $d\,J=1$, Eq.~\eqref{period} yields 
$t_\text{rev} \rightarrow \infty$, reflecting the fact that the mean-field equations
of motion do not predict any oscillations in this case -- in contrast to the
exact solution. With the absence of damping and the last observation, 
there are already two deficiencies of the mean-field approximation that we need to keep
in mind for the analysis that follows -- this makes the comparison to the exact solution 
an essential duty to ensure one has an idea of the limits of the validity of the
mean-field results presented in Sec.\ \ref{soft-core}.

\subsection{Exact vs mean-field results} \label{meanfieldB}

In equilibrium, a detailed comparison between the predictions of the mean-field theory
introduced before and exact quantum Monte Carlo simulations for the ground state of
hard-core bosons in the presence of a superlattice potential was presented in 
Refs.\ \cite{hen09,hen10}. The Gutzwiller approach was found to correctly predict 
the two phases present in the ground state of this model, namely, a superfluid 
phase for all fillings but $n=0,1/2,$ and 1 and for $n=1/2$ below a critical value of $A/J$
and a Mott insulator (a charge density wave) for $n=1/2$ above a critical value of $A/J$. 
However, Gutzwiller mean-field theory was shown to provide a poor estimate of
the critical value of $A/J$ for the superfluid-Mott-insulator transition. It overestimated
it by around 100\% in 2D and a 50\% in 3D.

As shown in the previous section, after the quench in the periodic system, the mean-field 
solution does not exhibit any damping for the amplitude of the oscillation whereas in the 
exact solution there obviously is damping. For this reason, we do not study the damping
any further. In the remainder of the article we therefore focus on 
the predictions of mean-field theory for the corrections to the revival time.

In order to be more quantitative, we define the relative deviation of the mean-field approximation
from the exact solution by
\begin{equation}\label{error}
\varepsilon(J) = \frac{\Delta t^\text{ex}_\text{rev}(J)-\Delta t^\text{mf}_\text{rev}(J)}
{\Delta t^\text{ex}_\text{rev}(J)}
\end{equation}
where $\Delta t^\text{ex}_\text{rev}(J)$ and $\Delta t^\text{mf}_\text{rev}(J)$
are the corrections to the revival time due to a finite value of $J$ for the exact 
and mean-field solutions, respectively.

\begin{figure}[!htb]
\begin{center}
\includegraphics[width=0.475\textwidth]
{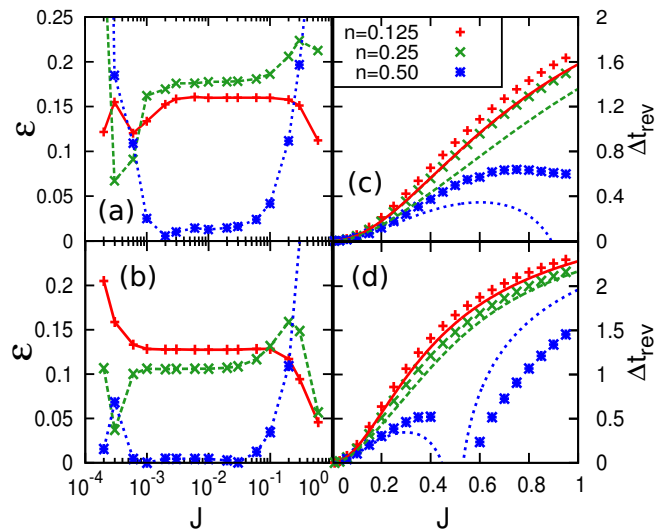}
\end{center} \vspace{-0.7cm}
\caption{(Color online) 
Comparison of the mean-field approximation and the exact solution
for 1D (a, c), and 2D (b, d).
In (a) and (b), we plot the error introduced in \eqref{error}
on a linear-log scale for the systems presented in Fig.\ \ref{exactHom}. 
In (c) and (d), we again show the results presented in 
Fig.\ \ref{exactHom} but this time on a linear scale together with the mean-field results,
where the latter are drawn as lines and were calculated via the analytical
solution \eqref{period}.
}
\label{compareMF}
\end{figure}

In Figs.~\ref{compareMF}(a) and \ref{compareMF}(b), we plot $\varepsilon(J)$ in a 
linear-log scale vs $J$. That figure shows an almost 
constant error over two decades ($10^{-3} \lesssim J \lesssim 10^{-1}$). For
$J \lesssim 10^{-3}$ rounding-off errors set in as one starts dealing with 
numbers $\sim 10^{-10}$; that is, the deviations seen in that region are not to be 
considered any further. The absolute value in the constant part of the deviation is around 
$16\%$ for 1D (a) and $11\%$ for 2D (b) for $n=0.125$ and $n=0.25$. At half filling,
interestingly, the deviation is yet much smaller. In all cases, it is obvious
that the mean-field approximation describes the 2D system better than the 1D case, 
even though the system size in 1D is one order of magnitude larger than in 2D
and mean-field is expected to be more accurate as the system size is increased.
 
In Figs.~\ref{compareMF}(c) and \ref{compareMF}(d), we present the same results
as in Fig.\ \ref{exactHom} and compare them to the mean-field predictions, but this time 
in a linear scale. This scale emphasizes the differences between the mean-field results
and the exact ones for values of $J$ close to $A=1$. Once again, it is obvious that 
the mean-field approximation works better in 2D than in 1D, and that it becomes a 
very good approximation of the exact results for $n=0.125$ and $n=0.25$.
We further note the already-mentioned case of $d J = 1$, which does not yield 
any revival in the mean-field approximation and therefore 
$\Delta t_\text{rev} \rightarrow \infty$ as the figures suggests. Interestingly,
there is a corresponding anomaly in the exact solution in 2D: there, in 
a neighborhood of $J=0.5$, the exact solution does not produce a symmetric
peak after the first revival oscillation, which makes it meaningless to determine a 
value for the revival time -- therefore these data points are missing. Finally,
Figs.\ \ref{compareMF}(a) and (b) show that at half filling mean-field theory provides 
the most accurate results for the quadratic region of small values of $J$, whereas
it provides the worse results for large values of $J$, as seen in (c) and (d).

\begin{figure}[!htb]
\begin{center}
\includegraphics[width=0.44\textwidth]
{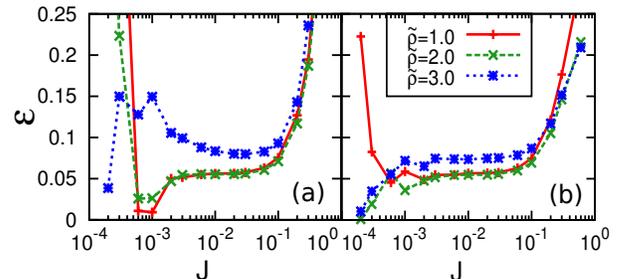}
\end{center} \vspace{-0.7cm}
\caption{(Color online) 
Plots of the error \eqref{error} for the 1D trapped case presented in Fig.\ \ref{exactTrap} 
for: (a) quench type (i) keeping the characteristic density constant and (b) quench type 
(ii) turning off the trap, both presented on a linear-log scale.}
\label{comparetrapMF}
\end{figure}

Results for $\varepsilon(J)$ in the 1D trapped 
case are presented in Fig.\ \ref{comparetrapMF} for the two types of quenches
introduced in the previous section: (i) the scenario of a constant characteristic
density and (ii) switching off the trapping potential. The behavior is 
qualitatively the same as just discussed for the homogeneous case
of Fig.\ \ref{compareMF}(a). Quantitatively, we find an error $\sim 5\%$, which is 
in between the values for low and half filling in Fig.\ \ref{compareMF}(a), and is similar
for both quench types. Such an intermediate value is expected because the trap causes
a density profile with different densities in different regions of the trap.

For the description of experiments, which are of more interest in 3D trapped geometries 
and very large system sizes, one can expect much smaller errors than the ones in
Fig.\ \ref{comparetrapMF}. Therefore, we conclude that the shift of the revival
time due to a finite hopping is correctly captured not only qualitatively but also 
quantitatively by the mean-field approximation described here. This is an interesting 
finding considering that, in contrast, the description of the evolution of the amplitude 
in terms of the mean-field approximation is incorrect even at the qualitative level.

\section{Results for soft-core bosons}\label{soft-core}

In Refs.\ \cite{jaksch98,bloch08} it has already been shown that ultracold bosonic gases 
in optical lattices can be well described by the bosonic Hubbard model \eqref{scb}.
In light of the recent results in Refs.\ \cite{will09,johnson09} mentioned in the 
introduction, we note that the interaction constant $U$ and the hopping 
amplitude $J$ in the Hamiltonian \eqref{scb} are the effective two-body interaction and 
one-body hopping, respectively, whose origin is not under discussion here. They may be 
obtained after multi-orbital renormalization effects are taken into account. Multi-orbital 
effects may also generate effective higher-body interactions that translate into 
additional frequencies during the collapse and revival of the matter-wave interference 
but are not considered here. These effects can be reduced by properly engineering the initial state. 
We focus on the effect that a 
finite effective hopping $J$ has on the fist revival of the matter wave.

Collapse and revival oscillations like the ones observed experimentally in Refs.\ \cite{greiner02,will09} 
have been reproduced in 1D bosonic \cite{kollath07} and fermionic 
\cite{manmana07} systems by means of numerically exact time-dependent renormalization
group techniques. Here we use mean-field theory to study 3D bosonic systems theoretically.
Following the results of the previous section, we expect the mean-field predictions 
for the tunneling-induced correction of the revival time of the value in the atomic limit $\Delta t_\text{rev}$ 
to be close to the exact results.

We proceed in the way introduced in the preceding section. For soft-core bosons, we solve 
the equations \eqref{jakschEq} for a cutoff of $n_\text{c}=7$ numerically. This 
cutoff ensures convergence with respect to the energy expectation value and the 
momentum distribution, and was successfully employed in Ref.\ \cite{zakr05} to study
the superfluid-Mott-insulator transition. Initially, the system is prepared in the 
Gutzwiller mean-field ground state of the trapped system at an intermediate interaction 
$U_\text{ini}=6J_\text{ini}$ ($J_\text{ini}=1$ sets our energy scale), which ensures 
the validity of a one-band model in experiments while the system is still far from the 
transition to the Mott insulator. Within the mean-field approximation 
$(U/J)_\text{crit}=34.8$ (for six nearest neighbors) \cite{fisher89,krauth92,sheshadri93}. 
At time $t=0$, the on-site interaction is doubled to $U_\text{fin}=12$ and we 
investigate the collapse and revivals for several values of $J_\text{fin}<1$ (where, as in the 
section about hard-core bosons, the notation $U_\text{fin}\equiv U$ 
and $J_\text{fin}\equiv J$ is used in all unambiguous cases). 

\begin{figure}[!htb]
\begin{center}
\includegraphics[width=0.41\textwidth]
{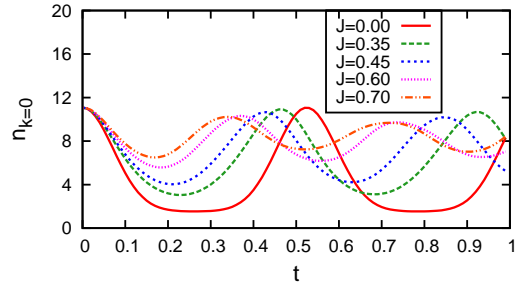}
\end{center} \vspace{-0.7cm}
\caption{(Color online) 
Dynamics of the zero momentum peak $n_{k=0}$ vs time $t$ for a
system with homogeneous potential and density $n=1.5$. Time is measured
in units of $\hbar/J_\text{ini}$.
These results are derived from the mean-field solution of the time evolution \eqref{jakschEq} 
for the Bose-Hubbard model after an interaction quench from $U_\text{ini}=6$ to $U_\text{fin}=12$. 
Several final values of the hopping constant $J$ are depicted.}
\label{SCBhomEvol}
\end{figure}

Figure \ref{SCBhomEvol} depicts the collapse and revival oscillations for the homogeneous case.
These predictions, within mean-field theory, correspond to the solution of a single site 
problem. They clearly exhibit a different functional 
form when compared to the ones for 
hard-core bosons in a superlattice potential Fig.\ \ref{exactHom}. Also, not only a
shift in the revival time is present but additionally a damping of the amplitude can 
be observed. We do not consider this damping effect, as we cannot make any statement
about the validity of the mean-field approximation for this quantity (see Sec.\ 
\ref{meanfield}). We note again that the revival time for the interaction quench
in the atomic limit is given by $t_\text{rev}=2\pi/U$.

\begin{figure}[!htb]
\begin{center}
\includegraphics[width=0.41\textwidth]
{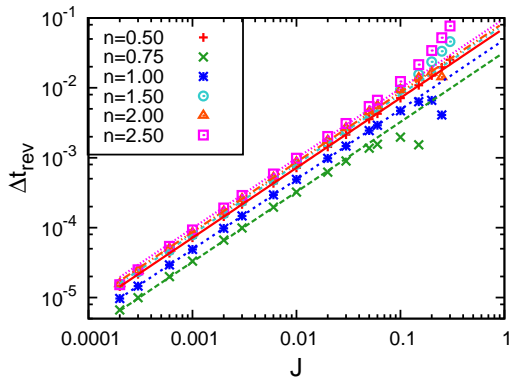}
\end{center} \vspace{-0.7cm}
\caption{(Color online) 
Plot of $\Delta t_\text{rev}$ vs $J$ for soft-core bosons in a homogeneous system 
and different densities after an interaction quench from 
$U_\text{ini}=6$ to $U_\text{fin}=12$. This is done for a single-site system,
as the homogeneous result is independent of the system size in the mean-field
approximation. Note that the results for different densities lie very close 
to each other. The linear fits are done for data points with $J\leq0.01$.}
\label{3DSCBhom}
\end{figure}

In Fig.~\ref{3DSCBhom}, we show $\Delta t_\text{rev}$ for several densities in the homogeneous 
case. We observe a linear relation emphasized by the fits for data points with $J\leq0.01$ 
in the figure. Since soft-core bosons are not subject to particle-hole symmetry, 
the behavior with increasing density is different from the one observed for hard-core bosons 
in Fig.~\ref{exactHom}. The smallest deviation from the atomic limit is no longer reached 
at half filling; instead, for soft-core bosons we find that it appears for a density $n\sim0.75$. 
Except for densities around this value, $\Delta t_\text{rev}$ is not strongly dependent on the density.
With respect to the dependence of $\Delta t_\text{rev}$ on both parameters $J$ and $U$,
we note that the same scaling as in the hard-core case
(with $A$ and $U$ interchanged) holds true: $t_\text{rev}(J,U)\equiv t_\text{rev}(J/U)/U$.

The calculations for the trapped case are more demanding computationally. This is because 
translational invariance is broken and one has to deal with all the lattice sites in the 
system. The results reported here are obtained for a lattice with $L=30\times30\times30=27000$ 
sites with $N=1000$ up to $N=11000$. As before, finite size effects for our observables of 
interest are extremely small. As a matter of fact, we found that it would be difficult to 
distinguish the results reported here from those of a $L=10^3=1000$ system. Again, 
results are presented for the two quench scenarios analyzed in detail in 
Sec.\ \ref{sec:hcbtrap} for hard-core bosons.

\begin{figure}[!htb]
\begin{center}
\includegraphics[width=0.38\textwidth]
{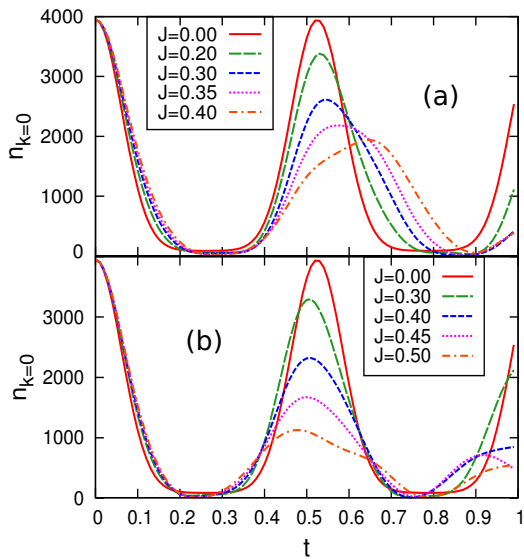}
\end{center} \vspace{-0.7cm}
\caption{(Color online) 
Plot of the time evolution of the momentum peak $n_{k=0}$ for $\sim11000$ soft-core bosons 
in a lattice with $L=30\times30\times30$ sites for: (a) quench scenario (i), and 
(b) quench scenario (ii) (see text). The initial state has a characteristic 
density of $\tilde\rho = 39.8$ and an on-site energy $U_\text{ini}=6$. At time $t=0$
the interaction is quenched to $U_\text{fin}=12$ and the curvature
of the trap is modified according to (i) and (ii).}
\label{SCBtrapEvol}
\end{figure}

Results for the time evolution of $n_{k=0}$ in the harmonic trap are shown in 
Fig.\ \ref{SCBtrapEvol} for an initial state with characteristic density
$\tilde\rho=39.8$ and the two different quench types: (i) keeping constant the 
characteristic density and (ii) turning of the trap. Here, we observe an effect that 
is qualitatively different from the the one seen in the hard-core limit 
(Fig.\ \ref{exactTrapEvol}) and the homogeneous soft-core case, namely, the revival time 
in the case with finite hopping exceeds the atomic value [Fig.\ \ref{SCBtrapEvol}(a)].
This effect is only observed in the quench scenario (i) for high characteristic densities. 
For quench type (ii) the effect is not present for any density [Fig.\ \ref{SCBtrapEvol}(b)].

\begin{figure}[!htb]
\begin{center}
\includegraphics[width=0.39\textwidth]
{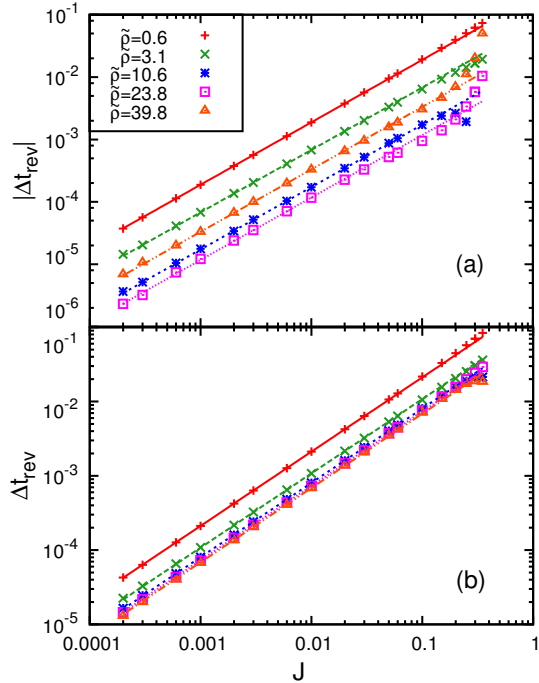}
\end{center} \vspace{-0.7cm}
\caption{(Color online)
Quantum corrections to the revival time ($\Delta t_\text{rev}$) vs $J$
for the scenarios: (a) keeping the characteristic density constant,
and (b) turning of the trap; and an interaction quench $U_\text{ini}=6 
\rightarrow U_\text{fin}=12$. Results are
presented for five different initial values of the characteristic density --
$\tilde{\rho}=0.6$, $\tilde{\rho}=3.1$, $\tilde{\rho}=10.6$,  $\tilde{\rho}=23.8$,
and  $\tilde{\rho}=39.8$ -- that correspond to the following values of the density in 
the middle of the trap: $n_\text{center}\sim0.5$, $n_\text{center}\sim 1.0$, 
$n_\text{center}\sim 1.5$, $n_\text{center}\sim 2.0$ and $n_\text{center}\sim2.5$.
Note that in (a) $\Delta t_\text{rev}$ has a negative sign for 
$\tilde{\rho}=23.8$ and $\tilde{\rho}=39.8$; therefore, the absolute 
value is depicted.
}
\label{SCBtrap}
\end{figure}

In Fig.\ \ref{SCBtrap}, $\Delta t_\text{rev}$ is plotted for several values of the
characteristic density of the initial state. In Fig.\ \ref{SCBtrap}(a), results are shown
for quench type (i). Note that we account for the fact that the revival time exceeds 
the atomic limit for high densities by plotting the absolute value of $\Delta t_\text{rev}$; 
in particular, $\tilde\rho=23.8$ and $\tilde\rho=39.8$ yield a negative value of 
$\Delta t_\text{rev}$. The strong dependence of the results on the characteristic density
of the initial system, or equivalently, the initial density in the center of the trap,
makes this scenario unsuitable for experimentally probing the values of $J$ after the 
quench. The experimental uncertainty of the filling in the center of the trap 
would lead to a large uncertainty in determining $J$.

Scenario (ii) seems to be a good candidate for the latter goal. As depicted in
Fig.\ \ref{SCBtrap}(b), for characteristic densities that are not too small 
($\tilde\rho \gtrsim 10$) i.e.\ for densities in the center of the trap that are 
$n_\text{center} \gtrsim 1.5$, we observe a weak dependence of the revival time on the 
characteristic density of the initial state. This is usually fulfilled in experiments like 
\cite{greiner02}. We also note that the relative deviation 
$\Delta t_\text{rev}^\prime = \Delta t_\text{rev}/t_\text{rev}^\text{atom}$
is only one order of magnitude smaller than the normalized hopping parameter
$J/U$, due to the linearity of the relation and a prefactor $\sim 0.1$.
This shows that the described effect is not small and one should be able to measure
it in experiments.

Finally, we should mention that we also performed calculations for different values of 
the interaction constant $U$ before and after the quench. They all exhibited a similar 
qualitative behavior as depicted in Fig.\ \ref{SCBtrap}. 
We therefore stress that our results do not depend on a particular value of $U$ but 
represent a general behavior that can be reproduced with experimentally relevant 
parameters. By comparing experimental results
and calculations within mean-field theory, plus using the expected experimental values
for $U$, one could then use experimentally measured values of $\Delta t_\text{rev}$ 
to determine the final value of $J$.

\section{Conclusion}\label{conclusions}

We have presented a detailed study of the dependence of collapse and revival 
properties of the matter-wave interference in lattice boson models on a finite tunneling 
amplitude after an interaction quench.

We first studied quenches of hard-core bosons on a superlattice potential. For those
systems, we presented exact numerical results in 1D and 2D, and compared
them with the approximated mean-field solution. Both approaches exhibited the same 
functional form of the correction to the revival time produced by finite hopping
parameters after the quench, with a leading behavior $\sim t^2/A^3$.  
The mean-field results were also shown to have, as expected, 
smaller errors in 2D as compared to 1D. Since the largest errors for homogeneous 2D systems were 
$\sim 10\%$ and for trapped 1D systems $\sim 5\%$ (in contrast to $\sim 17\%$ for the 1D homogeneous
case), we expect that in 3D trapped systems, mean-field theory should provide relatively
accurate results for the corrections to the revival time.

We then studied interaction quenches in the Bose-Hubbard model in 3D.
In this case, our analysis was solely based on the Gutzwiller mean-field theory.
We showed that for soft-core bosons the corrections to the revival time in the atomic limit, due
to finite values of $J$ after the quench, are $\sim J/U^2$. This is an effect that could 
be measured experimentally. Given the weak dependence of the correction on the initial 
density profile, provided the density in the center of the trap is greater than $n=1.5$, we 
have proposed that the corrections to the revival time measured experimentally could be 
used to determine the actual value of $J$ after the quench. The only input one would 
need is the experimental value of $U$ and the mean-field predictions from 
calculations similar to the ones presented here.

\begin{acknowledgments}
This work was supported by the US Office of Naval Research under Award No.\ N000140910966
and by the National Science Foundation under Grant No.\ OCI-0904597.
\end{acknowledgments}

\end{document}